# Bypassing damaged nervous tissue

## M. N. Shneider [1] and M. Pekker [2]

[1] Department of Mechanical and Aerospace Engineering, Princeton University, Princeton, NJ 08544, USA
[2] George Washington University, Washington, DC, 20052, USA

E-mails: m.n.shneider@gmail.com and pekkerm@gmail.com

**Abstract**

We show the principal ability of bypassing damaged demyelinated portions of nervous tissue, thereby restoring its normal function for the passage of action potentials. We carry out a theoretical analysis on the basis of the synchronization mechanism of action potential propagation along a bundle of neurons, proposed recently in [1]. And we discuss the feasibility of implement a bypass to restore damaged nervous tissue and creating an artificial neuron network.

**Introduction**

Violations of the nerve fiber integrity, such as partial demyelination or micro-ruptures, lead to malfunction of the nervous system and its components [2-6]. For example, a damage or loss of the myelin coating of neurons, which can be caused by various diseases, slows down or even blocks action potentials. This results in a variety of disorders, such as sensory impairment, multiple sclerosis, blurred vision, movement control difficulties, as well as problems with bodily functions and reactions [2-7]. Numerous articles and patents have been devoted to the problem of recovering damaged nervous tissue function [8-20].

Many of these proposed a recovery scheme of reading an electrical signal from one or several needle electrodes in contact with an individual or group of neurons, and then processing and transmitting it to another group of neurons with additional needle electrodes (Fig. 1).

In [21] recently showed a theoretical example of restoring the normal work of a partially demyelinated neuron cell, which remains alive and functional, by appropriate stimulation of the axon away from the damaged area that could lead to normal passage of the action potential bypassing the demyelinated area. Such stimulation can be a local change in membrane potential induced by the current in saline, similar to that in neuron synchronization [1].

In the first part of this paper, we will briefly discuss the physical assumptions of the model used in [1] for the correlation length evaluation of the propagation of action potentials in neighboring neurons. In the second part, we will consider the physical model of saltatory transition in myelin axons and neuron synchronization. In the third, we will apply the developed approach in part II for evaluating the correlation size of nearby neurons required to bypass damaged nervous tissue through non-contact electrodes. In the fourth, we will give an example of blocking the action potential propagation due to the demyelization of some part of the axon on the basis of the Goldman-Albus model of a toad neuron. In the fifth part we will consider the bypassing damaged nervous tissue by transfer of the action potential via "contactless" electrodes. In the Appendix, we will present the formulas and their derivations from [1] and this paper.

The purpose of this article is to demonstrate the theoretical possibility of bypassing damaged nerve fibers (or bypassing of a bunch of damaged nerve fibers) by using non-contact electrodes. At the present time, as noted above, the creation of such electrodes has become technically feasible. In order to better understand how practical is a bypass based on the non-contact electrodes, requires serious experimental studies.

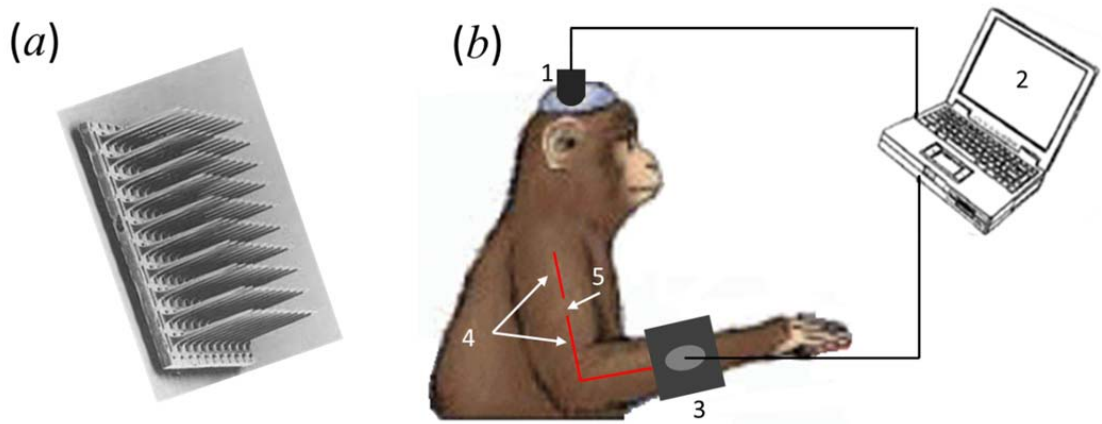

**Fig.1.** Restoration of nervous tissue functions using a bypass of needle electrodes. (a) – System of needle electrodes [20]. (b) Example of a bypass. 1 – system of needle electrodes for reading the signal from the motor areas of the brain; 2 –system for processing signal incoming from the motor areas of the brain and for generating an output for arm muscles control; 3 – system of needle electrodes for exciting neurons in the muscle tissue; 4 – nervous tissue connecting the motor areas of the brain to the muscles of the hand, 5 – rupture (damage) of the nervous tissue.

### I. Physical model of saltatory transition in myelin axons and neuron synchronization

It is known (see for example [22-24]) that excitation of the action potential in nerve fiber requires an application of the threshold stimulating 20-30mV voltage pulse of duration about 0.1ms to the non-myelinated section of the axon membrane. Moreover, it is sufficient to apply the voltage pulse to a small area of the initial part of the axon to stimulate the action potential in the entire fiber.

A mathematical model for calculating the fields and currents associated with the passage of the action potential in neurons has been developed in [25,26]. Subsequently, this theory has been used in [27, 28] for the calculation of the magnetic field outside an isolated axon. In particular, it was shown in [25, 26] that in the process of action potential propagation along a fiber, the change in potential on the outer surface of the membrane at the propagating action potential differs only by microvolts from the potential elsewhere on the outer surface of the membrane that is at rest.

Note that this is not a potential difference between the outer and inner surfaces of the membrane, but the potential difference $\Delta U$ on its outer surface (Fig. 2). It would seem that at such small variations in the surface potential the excitation conditions between neighboring neurons cannot be created, and thus for most mammalian nerve tissue the interaction of neurons is considered to be impossible [29-31].

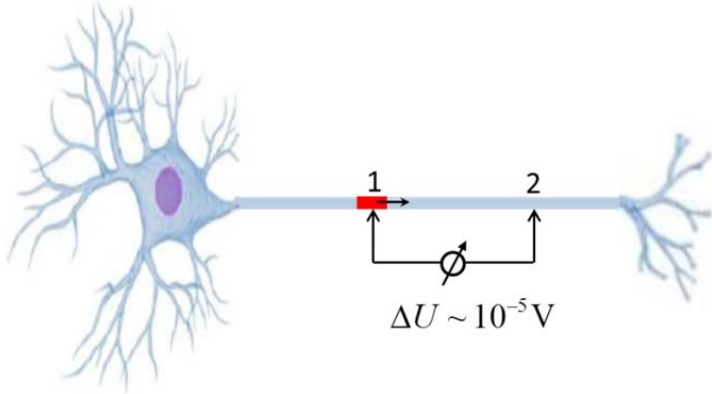

**Fig. 2.** Action potential propagation along a nerve fiber. 1 - The action potential, with the arrow indicating the direction of its propagation. 2 - Unperturbed area of the axon.

Recently in [1], we proposed a neighboring neuron synchronization mechanism (Ephaptic coupling), based on the fact that the action potential propagating along an axon is accompanied by currents flowing in the physiological extracellular electrolyte solution. We showed that these currents can cause a local charging of some area of the membrane of a neighboring neuron axon, where a potential difference arises that is sufficient to excite the action potential. This approach allows one to estimate the region of synchronization (the correlation scale) of action potentials in the cases of myelinated and non-myelinated fibers. This estimated correlation scale between the parallel initial segments of myelinated neurons is in agreement with experimental results [32]. The mechanism of synchronization of neurons proposed in [1] is similar to some extent to saltatory transition of action potentials between the nodes of Ranvier in myelinated axons [33-37]. Fig. 3 shows the interaction scheme of neurons through the currents induced in the interstitial fluid, considered in [1].

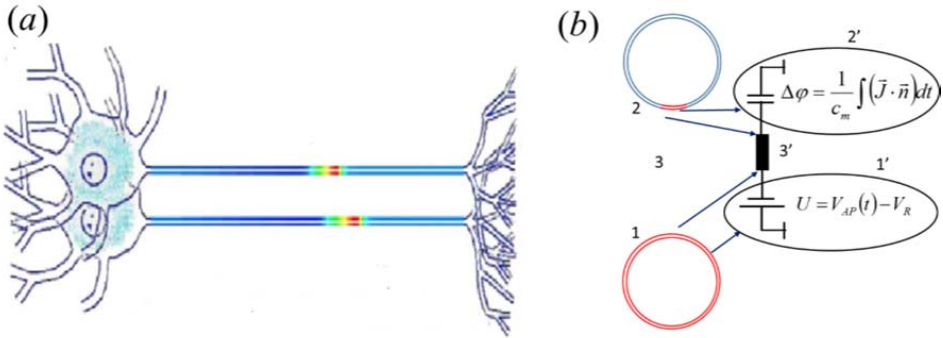

**Fig. 3.** Interaction scheme of neighboring neurons. (*a*) – Propagation of the action potential along the two coupled neurons. (*b*) – Excitation of the action potential by the primary (active) axon in the secondary (initially inactive) axon. 1 – primary axon; 2 – secondary axon: red color shows the area that is charged by volume currents induced in the extracellular electrolyte 3; 1', 2', 3' – equivalent electrical scheme of charging of the membrane of the secondary axon. $c_m$ is the capacity of the axon membrane per unit area, $\vec{J}$ is the local current, that charges the membrane of secondary axon, $\vec{n}$ is the unit vector normal to the surface of the membrane, $\Delta\varphi$ is the local potential difference between the capacitor plates (outer and inner surfaces of the primary axon membrane), $U$ is the effective EMF source (potential difference between the surfaces of the membrane of the axon, along which an action potential propagates), and $V_{AP}$ and $V_R$ are the action and resting potentials respectively.

## II. Potentials and currents in the vicinity of a neuron at action potentials propagation

The fluid inside and outside a neuron is an electrolyte with relatively high conductivity: $\sigma \sim 1-3\ \Omega^{-1}\mathrm{m}^{-1}$ [38]. Therefore, the relaxation time of the volume charge perturbations in it is about the Maxwellian time $\tau_M = \varepsilon\varepsilon_0/\sigma \sim 10^{-9}\,\mathrm{s}$ ($\varepsilon_0$ is the permittivity of vacuum, $\varepsilon \approx 80$ is the dielectric constant of water) [39], which is several orders of magnitude less than the characteristic time of excitation and relaxation of the action potential in the axons.

The size of the quasi-neutrality violations in the electrolyte is determined by the Debye length:

$$\lambda_D = \sqrt{\varepsilon\varepsilon_0 k_B T \Big/ \left(\sum_{j=1}^{J} n_j q_j^2\right)} \tag{1}$$

Here $k_B$ is the Boltzmann constant, $T$ is the temperature, and $n_j$ and $q_j$ are the density and the charge of the ions. The density of the ions in the electrolyte outside and inside the cells is close to the the ion density in saline: $n_j \approx 2\cdot 10^{26}\ \mathrm{m}^{-3}$ [40]. For typical parameters of interstitial fluid at $T \approx 300$ K, equation (1) yields $\lambda_D \approx 0.5$ nm. That is, $\lambda_D$ is at least four orders of magnitude smaller than the typical axon radius $R_0 \sim 3-10\ \mu\mathrm{m}$.

This means that the effects associated with the violation of quasi-neutrality in the fluid inside and outside the axon, at the action potential propagation, can be neglected. Accordingly, one can use the current continuity equation in order to find the potential distribution inside and outside the axon:

$$\nabla \cdot \vec{J} = 0,\ \vec{J} = -\sigma\nabla\varphi. \tag{2}$$

Since the conductivity $\sigma$ in the electrolyte is a constant, equation (2) can be written as

$$\Delta\varphi = 0, \tag{3}$$

with the Neumann boundary conditions on the surface of the membrane:

$$\left.\frac{\partial\varphi}{\partial r}\right|_{r=R_0} = -\frac{J_m}{\sigma}, \tag{4}$$

and the Dirichlet condition, away from the surface of the membrane:

$$\varphi_{r\to\infty} = 0. \tag{5}$$

Here $J_m$ is the density of the total radial current flowing through the membrane of the axon, which is the sum of the ionic $J_{ion}$ and capacitive $J_{capas} = C_m \partial U/\partial t$ current densities, in which $C_m \approx 2\cdot 10^{-6}\,[\mathrm{F/cm}^2]$ is the capacity per unit of the membrane surface, $U = V_{AP} - V_R$ is the potential difference between the surfaces of the membrane of the axon, along which an action potential $V_{AP}$ propagates, and $V_R$ is the resting potential (Fig. 3). In the charging process of the myelin section of the axon, the ion current component flowing through the membrane can be neglected, because the size of the node of Ranvier (where the ion current is non-zero) $L_R \approx 1-2.5\ \mu\mathrm{m}$ is much smaller than the characteristic size of the myelinated section, $L_M \sim 100-2000\ \mu\mathrm{m}$ [7] (Fig. 4).

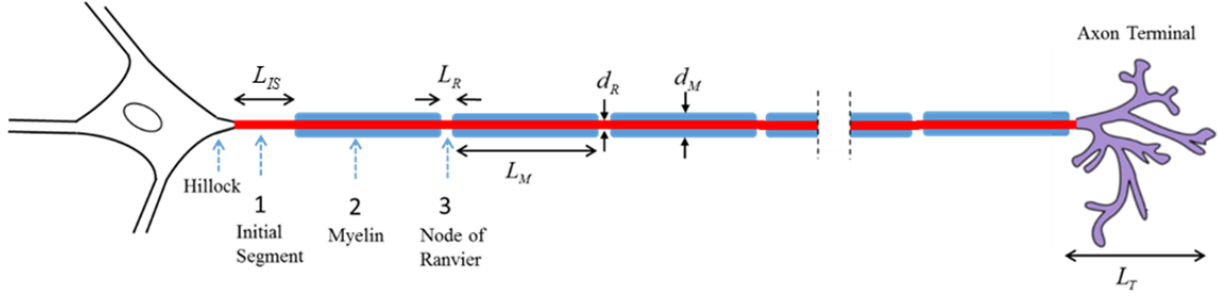

**Fig.4.** A neuron with the characteristic elements of the myelinated axon. Typically, the size of $L_{IS}$ is in the range of 20-50 μm, but it can reach 75 μm and even 200 μm, $L_R \approx 1-2.5\,\mu m$, $L_M \sim 100-2000\,\mu m$, and $L_T \sim 10-2000\,\mu m$ [22].

Following [25, 26], we assume that the axon membrane is a thin-walled cylindrical capacitor. In this case, a good simplified assumption is that a traveling wave of the radial current on the surface of the axon propagating along it with the action potential velocity $v$, has the Gaussian dependence

$$J_m = J_0 \exp\left(-\left(\frac{z+vt}{L}\right)^2\right), \tag{6}$$

and the potential $\varphi^e$ and currents $J_z^e$ and $J_r^e$ outside the axon ($r \geq R_0$) have the form (see Appendix):

$$\varphi^e(r,z,t) = \frac{J_0 L}{\sqrt{\pi}\sigma_e} \int_0^{+\infty} \exp\left(-\left(\frac{L\xi}{2R_0}\right)^2\right) \frac{K_0\left(\xi \frac{r}{R_0}\right)}{\xi K_1(\xi)} \cos\left(\xi \frac{(z+vt)}{R_0}\right) d\xi, \tag{7}$$

$$J_z^e(r,z,t) = \frac{J_0 L}{\sqrt{\pi} R_0} \int_0^{+\infty} \exp\left(-\left(\frac{L\xi}{2R_0}\right)^2\right) \frac{K_0\left(\xi \frac{r}{R_0}\right)}{K_1(\xi)} \sin\left(\xi \frac{z+vt}{R_0}\right) d\xi, \tag{8}$$

$$J_r^e(r,z,t) = \frac{J_0 L}{\sqrt{\pi} R_0} \int_0^{+\infty} \exp\left(-\left(\frac{\xi L}{2R_0}\right)^2\right) \frac{K_1\left(\xi \frac{r}{R_0}\right)}{K_1(\xi)} \cos\left(\xi \frac{z+vt}{R_0}\right) d\xi. \tag{9}$$

In equations (7-9), $K_0$ and $K_1$ are the modified Bessel functions of the second kind of zero and first order, and $\sigma_e$ is the conductivity of the extracellular electrolyte. We present the detailed calculation of the formulas for the potential and currents inside the cylindrical capacitor filled with a conductive fluid in the Appendix.

The potential difference on the membrane of the secondary axon (see Fig. 3), without taking into account charge leakage, is determined by the charges on the membrane surfaces. Knowing the time dependence of the charging current density and its distribution in space, one can calculate the potential difference across any area of the membrane of the initially inactive secondary axon, depending on the position of this area:

$$\Delta U(z) = \frac{1}{C_m} \int_0^\infty (\vec{J} \cdot \vec{n}) dt. \tag{10}$$

## III. The correlation lengths for the action potentials excitation in the case of myelinated axons

In [1], we gave numerical estimations of the correlation scales (required for the initiation of the action potential in neighboring neurons) for the cases of the myelinated and non-myelinated (squid) axons. There, we approximated the radial current on the surface of myelinated and non-myelinated axon membranes by a single exponential function (6), which does not describe the discharging of the membrane surface of a secondary axon (see, Fig. 3) after the passage of the action potential along the primary axon, because the currents in the form (8) and (9) are always positive. Hence, here we give a more accurate estimate of the correlation scale for the case of myelinated fibers. As in [36,41,1], we assume, for definiteness that $v = 18.4$ m/s, and $R_0 = 4.5$ μm.

Fig. 5 shows the dependencies of the action potential at the arbitrary point $z = 0$ on time, and in the time moment $t = 0$ on z. Curve 1 corresponds to the action potential taken from [36,41,1], and 2 to the approximation formula for current on the membrane surface of the axon:

$$J_m = J_0\left(e^{-(z+vt)^2/L^2} - \frac{L}{L_1}e^{-(z+z_0+vt)^2/L_1^2}\right), \qquad (11)$$

where $J_0 = 0.0024$ A/cm$^2$, $L = 0.0011$ m, $L_1 = 0.0088$ m, and $z_0 = 0.0099$ m.

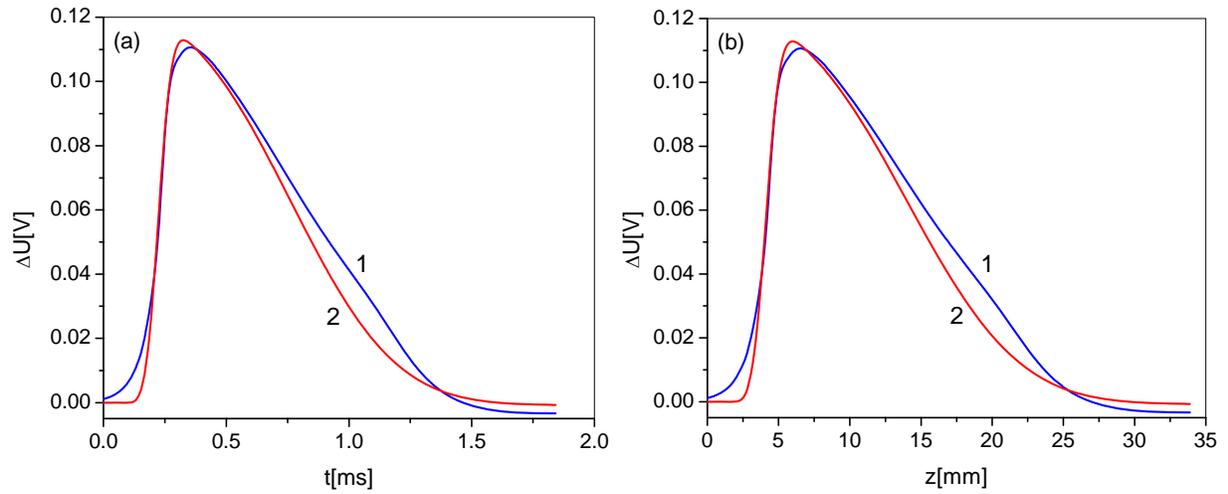

**Fig. 5.** Dependencies of the action potential at the arbitrary point $z = 0$ on time (a), and at a chosen time moment $t = 0$ on z (b). Curve 1 corresponds to the action potential taken from [36, 41, 1], and 2 to the approximation formula (11) for the current on the surface of the axon membrane. The value of $\Delta U$ in plot (a) is given by (10), while in plot (b) by the formula $\Delta U(z) = \frac{1}{C_m}\int_0^{z/v} J_m(t')dt'$, with $C_m = 10^{-6}$ F/cm$^2$.

Fig. 6 shows the dependence of the potential on the membrane of the secondary axon as a result of charging by current induced by the primary axon. As the action potential excitation in the axon occurs at $\Delta U \approx 0.02 - 0.025$ V, and the voltage pulse duration is of the order of 0.1 ms, it follows from Fig. 6 that the action potential excitation in the neighboring neuron is possible if the distance between the axons of primary and secondary neurons does not exceed 18 μm μm. In other words, the correlation scale for myelinated fibers with parameters considered above is $18\,\mu m$.

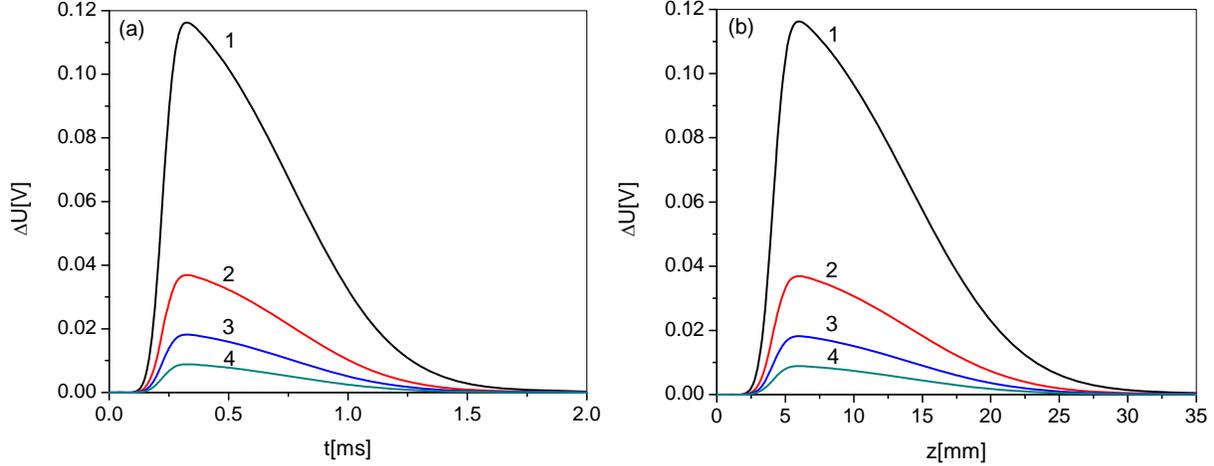

**Fig. 6.** Dependencies of the potential on the secondary axon membrane due to the charging by the currents induced by the primary axon. (a) – dependence on time at the point $z = 0$; (b) – on $z$ at $t = 0$. Curve 1 corresponds to the touch between the axons of secondary and primary neurons, curve 2 corresponds to the distance between the primary and secondary axons (see Fig. 1) equal to $R - R_0 = 9.68$ μm, curve 3 to $R - R_0 = 24.25$ μm, and curve 4 to $R - R_0 = 54.45$ μm. In all cases, $C_m = 10^{-6}$ F/cm$^2$.

## IV. Demyelinated nervous fiber

Violation of the nervous tissue activity occurs either due to rupture of the nerve fiber, or blocking of the action potential propagation along the axons due to thinning of the myelin layers, increasing of the transmembrane leakage conductance, and growth of isolating coatings over the nodes of Ranvier [2-7].

An example of blocking of the action potential propagation due to the demyelization of some part of the axon on the basis of the Goldman-Albus model of a toad neuron [36] was considered in [41]. In this model, which has been used by many authors and reproduced in [41], a myelinated nerve fiber is represented as a leaky transmission line with uniform internode sections with a length $L_M$ and an outer diameter $d_M$ (myelin sheath inclusive) separated by nodes of Ranvier with a length $L_R$ and outer diameter $d_R$, which is equal to the diameter of an axon inside the myelin coating.

The voltage $U(x,t)$ across the membrane in the internode region is found as a solution to the equation of potential diffusion given in [36]:

$$\frac{\partial U}{\partial t} = a \frac{\partial^2 U}{\partial z^2} + bU, \tag{12}$$

where $a = 1/R_1 C_1$ and $b = -1/R_m C_1$. Here, $C_1 = k_1 / \ln(d_M / d_R)$ and $R_1 = \dfrac{R_i}{\pi (d_R / 2)^2}$ are the myelin capacitance and resistance per unit length, $R_m = k_2 \ln(d_M / d_R)$ is the myelin resistance times unit length, $k_1$ and $k_2$ are constants, and $R_i$ is the axoplasm specific resistance. All parameters and constants are the same as in [36]. The diameters of the myelinated sections of the axon and nodes of Ranvier are $d_M = 15$ and $d_R = 9$ μm respectively.

The action potential in the nodes of Ranvier $U_{AP}(t)$, with the corresponding resting potential $V_R = -70$ mV, is defined by the Frankenhaeuser–Huxley equation [42].

Fig. 7a shows an example of the action potential propagation along an undamaged myelinated axon, calculated in the framework of approximations and data used in [36]. For definiteness, we assumed that demyelination occurs between 12 and 22 nodes of Ranvier in the calculations of the damaged axon model. We mimicked neuron demyelination by assuming a myelin sheath of a smaller thickness: the diameter of the myelinated section in demyelinated areas was $d_M = 1.025$ and $d_R = 9.225$ μm. That is, in demyelinated areas the capacitance per unit length increases sharply, and the resistance decreases. Fig. 7b shows the computed results of the action potential in the damaged demyelinated axon. Under these assumed conditions, demyelination causes complete blockage of the action potential. However, as was shown in [24], that a rise of the resting potential at the first node of Ranvier after the demyelinated area by 38 mV (~ 31.6% of the maximum amplitude of the action potential) (Fig. 7c) is enough to restore normal propagation of the action potential (Fig. 7c).

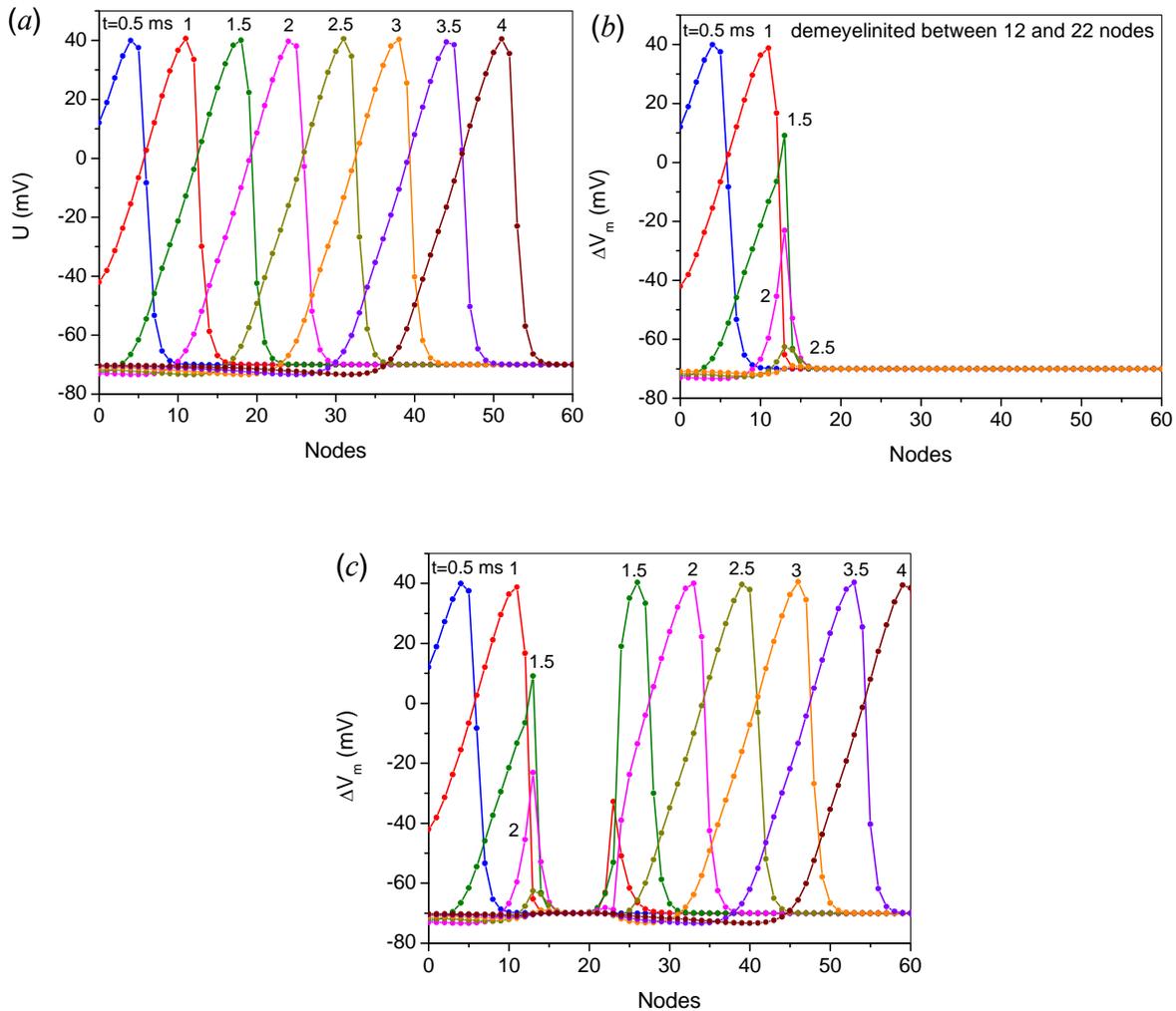

**Fig. 7.** (a) – Saltatory propagation of action potential in normal myelin fibers. (b) – Blocking of the action potential in a demyelinated area. (c) – Bypass of the demyelinated area (between 12 and 22 nodes of Ranvier).

A simple mathematical model of myelin fibers charging by the action potential excited in the nodes of Ranvier [1]. Fig. 8 shows the dependence of the maximal voltage drop across the extended myelinated section on the distance from the node of Ranvier (on the left at z = 0), where the action potential was excited. If the length of the myelin segment is greater than 4 mm, then the voltage does not reach the threshold (~ 38 mV) at the next node of Ranvier, and consequently the action potential is blocked. In fact, the length of the myelinated segments for the considered toad axons ≈ 2 mm, which ensures normal saltatory propagation of the action potential. Fig. 8b also suggests that an overgrowth of an isolating coating by only one node of Ranvier is sufficient to block the action potential propagation.

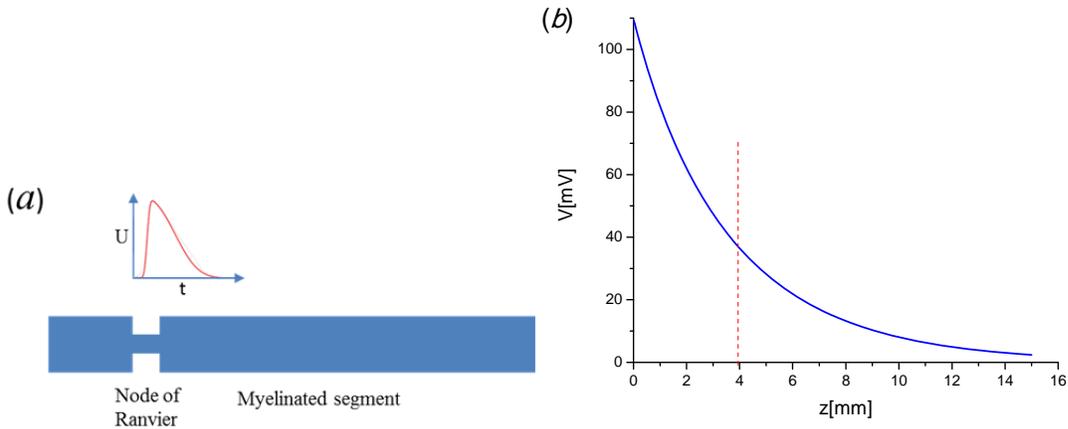

**Fig. 8.** (a) Simplified sketch of a part of a myelinated axon, including the active node of Ranvier, where the action potential $U(t) = V_{AP}(t) - V_R$, is initiated, that charges the adjacent semi-infinite myelinated segment, which is an extended cylindrical capacitor with capacitance per unit length $C_1$. (b) Dependence of the maximum voltage across the membrane of the myelinated segment on the distance from the active node of Ranvier, located at z = 0.

### V. Bypassing damaged nervous tissue by transfer of the action potential via "contactless" electrodes

All the axon damages mentioned earlier, related to demyelization and overgrowth of the nodes of Ranvier with an isolating coating, are not accompanied by the death of neurons. Therefore, a natural question arises: whether it is possible to bypass a damaged area of nerve tissue, and thus to maintain its normal functioning. Fig. 9 shows an example of bypass realization that solves this problem. During action potential propagation, the potentials of electrodes 1 and 2 are different; therefore, an electric connection occurs between areas A and B: part of the charge from region A flows into region B, which automatically leads to charging of the membrane of the axon in B. If the transferred charge is large enough that the local potential difference across the membrane of the nodes of Ranvier in area B exceeds the threshold value ~ 20-30 mV, an action potential is excited in B. Note the interesting possibility of increasing the efficiency of the bypass. Since the charge flowing from area A to B persists, by reducing the size of electrode 2 by N times compared to electrode 1, the current density in area B, as well as the efficiency of charging the membrane of the axon in B, increases by N times. The efficiency of the bypass can also be increased by placing electrode 2 closer to the membrane of the axon.

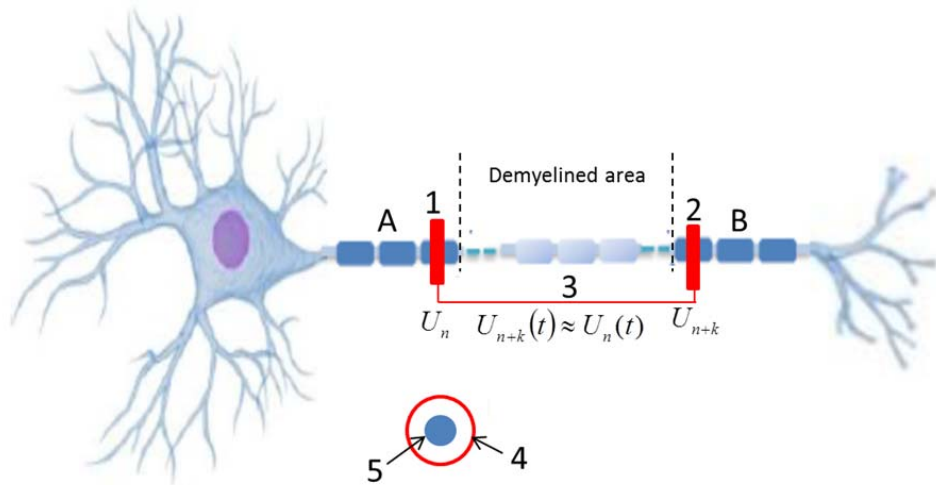

**Fig. 9.** Schematic view of damaged myelin fibers showing presence of a demyelinated area, in which the nodes of Ranvier are absent (overgrown). A and B are regions of the axon that are not damaged. 1 and 2 are the contactless electrodes: electrode 1 accumulates the charge induced by the propagating action potential, and electrode 2 is located beyond the damaged section of the myelinated fiber. 3 is an insulated conductor connecting the two electrodes. And 4 and 5 are the contactless electrode and undamaged section of the myelinated fiber in a cross-section view.

When the nerve fiber is fully damaged, the action potential can be transmitted to nearby, or even to a sufficiently remote neuron, via contactless electrodes connected by the insulated conductor, as shown in Fig. 10.

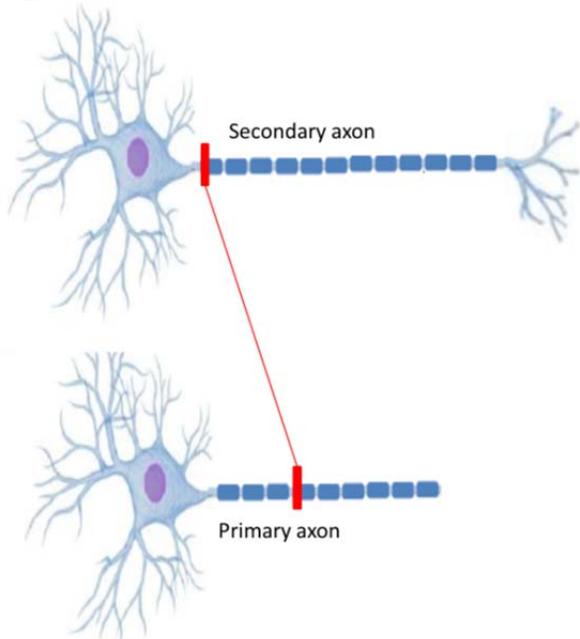

**Fig. 10.** Remote connection between two neurons by means of contactless extended electrodes.

The expression for the charge transmitted from area A to area B by the contactless ring electrode system 1-2, shown in Fig. 9, is

$$Q(t) = 2\pi R \int_0^l \int_{-\infty}^t J_r^e(R,z,t)dtdz, \qquad (13)$$

where $R$ and $l$ are the radius and length of the ring electrode respectively, and $J_r^e$ is the radial current induced in the extracellular electrolyte by the action potential. We give an explicit formula for $Q(t)$ in the Appendix for the case of Gaussian distributions of radial current (6) on the membrane surface.

Fig. 11 shows the dependencies of the charge transmitted in the electrode system 1-2 between regions A and B on the electrode length $l$ and radius $R_{coil}$. The current on the membrane surface has the form (11) with the same parameters $J_0$, $L$, $L_1$, $z_0$, and $R_0$ as in the calculations presented in part 2.

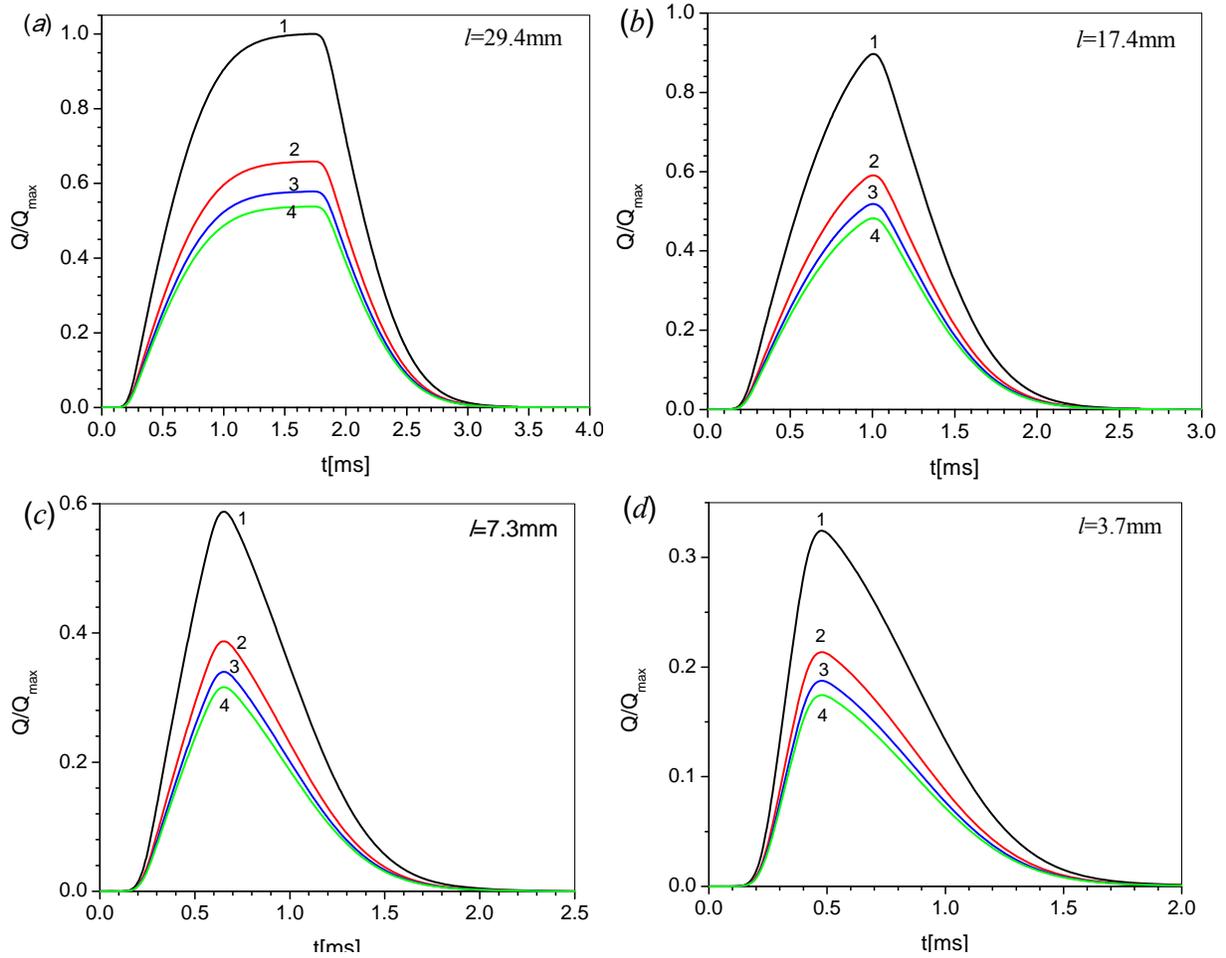

**Fig. 11.** Dependence of the charge collected by the ring electrode on time normalized to that of the maximal charge. In (a) the length of the ring is equal to 29.4 mm, in (b) to 14.7 mm, in (c) to 7.3 mm, and in (d) to 3.7mm. Curve 1 corresponds to $R_{coil} = R_0 = 4.5$ μm, curve 2 to $R_{coil} = 13.18$ μm, 3 to $R_{coil} = 28.75$ μm, and 4 to $R_{coil} = 58.95$ μm.

In this paper, we have given only two examples of using "contactless" electrodes to transmit the action potential, bypassing the damaged areas in a single axon, or between individual neurons. In fact, based on a similar approach, it is possible to construct a neural network with a correlation scale, significantly greater than the natural one.

We have shown that it is possible to bypass damaged demyelinated portions of nervous tissue, thus restoring its normal function. All of this is true not only for a single axon, but also for a neuron ensemble, as in the spinal cord, for example. The results presented in this paper, require experimental verification to evaluate the possibility of its implementation in biomedical practice. To observe the propagation, complete blockage or bypass of the action potential may be possible by detecting the second harmonic generation on a sample of nerve tissue, as suggested in [41,43]. These results demonstrate that the second harmonic response can serve as a local probe for the state of the myelin sheath, providing a high contrast detection of neuron demyelination.

It should be noted that it is also possible to block the action potential in an undamaged nerve fiber, leading to reversible anesthesia, while maintaining negative voltage below the threshold in some intermediate sections of the axon at the nodes of Ranvier.

### Conclusions

1. A mathematical model for the action potential saltatory propagation in myelin axons was developed.

2. A scale of correlation for the nearby neurons which is required to bypass the damaged nervous tissue through non-contact electrodes was calculated on the basis of the mathematical model for the action potential saltatory propagation.

3. An example of blocking of the action potential propagation due to the demyelization of some part of the axon on the basis of the Goldman-Albus model of a toad neuron is given.

4. Note, it is not necessary that non-contact electrodes considered in the work should be only in a ring form, they may be made as a semi-rings, small plates, or grids. The main requirement is that they effectively collect the current in the vicinity of an axon initiated by the action potential.

5. We have considered bypassing of the damaged areas for a single neuron. However, the principle of using of non-contact electrodes remains the same also for the aggregate (bunch) of simultaneously working neurons.

6. A positive feature of non-contact electrodes is the fact that for bypassing of the damaged section of the nerve fibers, there is no need to use an external electronic device to amplify the signal from the active fiber section. However, if there is a need for filtering, amplification, delaying or suppressing of the signal from the active fiber or the bunch of fibers, the scheme proposed in the paper for communication of one fiber to another allows to do this by using conventional electronic schemes, designed for system of needle electrodes [8-15].

7. In the past 10-15 years it has created a completely new technology for creating electrodes of submicron size (from 20 to 200 nanometers). These electrodes may be made in the form of needles, thin wires or grid. Therefore, the creation of the required extended electrodes of the order of tens or hundreds $\mu m^2$ and with thickness of a few $\mu m$ in the form of rings, small plates or grid is not a technical problem today [19,20,45-47].

### Appendix

In accordance with [25, 26], we assume the myelinated section of the axon is an endless thin-walled cylindrical capacitor with an external radius $R_0$, which is placed into a conductive liquid with the conductivities inside and outside the cylinder, $\sigma_e$ and $\sigma_i$, respectively. The action potential propagates along the axon, which initiates radial surface currents $J_m = C_m \partial U / \partial t$. In this case, the potential and currents outside of the axon ($r \geq R_0$) are given by formulas [25,26]:

$$\varphi^e(r,z,t) = \frac{1}{2\pi} \int_{-\infty}^{+\infty} \varphi_k^e(t,R_0) \frac{K_0(|k|\cdot r)}{K_0(|k|\cdot R_0)} e^{-ik\cdot z} dk \tag{A.1}$$

$$J_z^e(r,z) = \sigma_e E_z = -\sigma_e \frac{\partial \varphi}{\partial z} = \frac{i\sigma_e}{2\pi} \int_{-\infty}^{+\infty} \varphi_k^e(t,R_0) \frac{K_0(|k|\cdot r)}{K_0(|k|\cdot R_0)} k e^{-ik\cdot z} dk \tag{A.2}$$

$$J_r^e(r,z) = -\sigma_e \frac{\partial \varphi}{\partial r} = \frac{\sigma_e}{2\pi} \int_{-\infty}^{+\infty} \varphi_k^e(t,R_0) \cdot \frac{|k| K_1(|k|\cdot r)}{K_0(|k|\cdot R_0)} e^{-ik\cdot z} dk, \tag{A.3}$$

in which $K_0$ and $K_1$ are the modified Bessel functions of the second kind of zero and first order, and $\varphi_k^e(t,R_0)$ is the potential on the outer surface of the membrane.

Let the current $J_m = J_r(R_0,z,t)$ at the membrane surface be given by

$$J_m = \frac{1}{2\pi} \int_{-\infty}^{\infty} J_{m,k} e^{-ikz} dk = \frac{1}{2\pi} \int_{-\infty}^{\infty} J_{r,k}^e(R_0,t) e^{-ikz} dk. \tag{A.4}$$

Equating (A.3) to (A.4) at $r = R_0$, we obtain

$$J_{m,k}^e(t) = J_{r,k}^e(R_0,t) = \sigma_e \frac{|k| K_1(|k|R_0)}{K_0(|k|R_0)} \varphi_k^e(t,R_0). \tag{A.5}$$

Accordingly,

$$\varphi_k^e(t,R_0) = \frac{1}{\sigma_e} \frac{K_0(|k|R_0)}{|k|\cdot K_1(|k|R_0)} J_{m,k}(t). \tag{A.6}$$

This implies that

$$\varphi^e(r,z,t) = \frac{1}{2\pi\sigma_e} \int_{-\infty}^{+\infty} \frac{K_0(|k|r)}{|k| K_1(|k|R_0)} J_{m,k}(t) e^{-ik\cdot z} dk \tag{A.7}$$

$$J_z^e(r,z) = \frac{i}{2\pi} \int_{-\infty}^{+\infty} \frac{K_0(|k|r)}{K_1(|k|R_0)} \frac{k}{|k|} J_{m,k}(t) e^{-ik\cdot z} dk \tag{A.8}$$

$$J_r^e(r,z) = \frac{1}{2\pi} \int_0^{+\infty} \frac{K_1(|k|r)}{K_1(|k|R_0)} J_{m,k}(t) e^{-ik\cdot z} dk. \tag{A.9}$$

If the current on the surface of the axon is a traveling wave with a Gaussian profile:

$$J_m = J_0 \exp\left(-\left(\frac{z+vt}{L}\right)^2\right),\tag{A.10}$$

then its Fourier transform $J_{m,k}$ is given by:

$$J_{m,k} = \sqrt{\pi}LJ_0 \exp\left(-L^2k^2/4 - ikvt\right).\tag{A.11}$$

Using (A.11), formulas (A.7) - (A.9) become

$$\varphi^e(r,z,t) = \frac{LJ_0}{\sqrt{\pi}\sigma_e} \int_0^{+\infty} \exp\left(-\left(\frac{L\xi}{2R_0}\right)^2\right) \frac{K_0\left(\xi \frac{r}{R_0}\right)}{\xi K_1(\xi)} \cos\left(\xi\frac{(z+vt)}{R_0}\right)d\xi \tag{A.12}$$

$$J_z^e(r,z,t) = \frac{J_0 L}{\sqrt{\pi}R_0} \int_0^{+\infty} \exp\left(-\left(\frac{L\xi}{2R_0}\right)^2\right) \frac{K_0\left(\xi \frac{r}{R_0}\right)}{K_1(\xi)} \sin\left(\xi\frac{z+vt}{R_0}\right)d\xi \tag{A.13}$$

$$J_r^e(r,z,t) = \frac{J_0 L}{\sqrt{\pi}R_0} \int_0^{+\infty} \exp\left(-\left(\frac{\xi L}{2R_0}\right)^2\right) \frac{K_1\left(\xi \frac{r}{R_0}\right)}{K_1(\xi)} \cos\left(\xi\frac{z+vt}{R_0}\right)d\xi .\tag{A.14}$$

The charge on the surface of the ring electrode of radius $R$ and length $l$, excluding the feedback effect of the potential on the current distribution, is equal to:

$$Q(t) = 2\pi R \int_0^l \int_{-\infty}^t J_r^e(R,z,t)dtdz =$$

$$2\pi R \frac{J_0 L}{\sqrt{\pi}R_0} \int_0^l \int_{-\infty}^t \int_0^{+\infty} \exp\left(-\left(\frac{L\xi}{2R_0}\right)^2\right) \frac{K_0\left(\xi \frac{r}{R_0}\right)}{K_1(\xi)} \cos\left(\xi\frac{z+vt'}{R_0}\right)d\xi dt' dz = \tag{A.15}$$

$$\frac{2\pi R}{v} \frac{J_0 L}{\sqrt{\pi}} \int_0^l dz \int_{-\infty}^{(z+vt)/R_0} \int_0^{+\infty} \exp\left(-\left(\frac{L\xi}{2R_0}\right)^2\right) \frac{K_1\left(\xi \frac{r}{R_0}\right)}{K_1(\xi)} \cos(\xi\eta)d\xi d\eta$$

.

At $t \to \infty$,

$$Q = \frac{2\pi Rl}{v} \frac{J_0 L}{\sqrt{\pi}} \int_{-\infty}^{\infty} \int_0^{+\infty} \exp\left(-\left(\frac{L\xi}{2R_0}\right)^2\right) \cdot \frac{K_1\left(\xi \frac{r}{R_0}\right)}{K_1(\xi)} \cos(\xi\eta)d\xi d\eta .\tag{A.16}$$

Carrying out calculations similar to (A.1) - (A.19), the formulas for the potential and current inside the axon as functions of the radial currents on the membrane cylindrical surface become

$$\varphi^i(r,z,t) = -\frac{1}{2\pi\sigma_i} \int_{-\infty}^{+\infty} J_{r,k}^i(R_0,t) \frac{I_0(|k|\cdot r)}{|k|I_1(|k|\cdot R_0)} e^{-ikz} dk \qquad (A.17)$$

$$J_z^i(r,z) = -\frac{1}{2\pi} \int_{-\infty}^{+\infty} J_{r,k}^i(R_0,t) \frac{I_0(|k|\cdot r)}{I_1(|k|\cdot R_0)} \frac{ik}{|k|} \cdot e^{-ik\cdot z} dk \qquad (A.18)$$

$$J_r^i(r,z) = \frac{1}{2\pi} \int_{-\infty}^{+\infty} J_{r,k}^i(R_0,t) \frac{I_1(|k|\cdot r)}{I_1(|k|\cdot R_0)} e^{-ik\cdot z} dk \quad . \qquad (A.19)$$

In the case of the Gaussian distribution of the longitudinal currents on the membrane cylindrical surface (A.10), we get

$$\varphi^i(r,z,t) = -\frac{LJ_0}{\sqrt{\pi}\sigma_i} \int_0^{+\infty} \exp\left(-\left(\frac{L\xi}{2R_0}\right)^2\right) \frac{I_0\left(\xi\frac{r}{R_0}\right)}{\xi I_1(\xi)} \cos\left(\xi\frac{(z+vt)}{R_0}\right) d\xi \qquad (A.20)$$

$$J_z^e(r,z,t) = -\frac{J_0 L}{\sqrt{\pi}R_0} \int_0^{+\infty} \exp\left(-\left(\frac{L\xi}{2R_0}\right)^2\right) \frac{I_0\left(\xi\frac{r}{R_0}\right)}{I_1(\xi)} \sin\left(\xi\frac{z+vt}{R_0}\right) d\xi \qquad (A.21)$$

$$J_r^e(r,z,t) = \frac{J_0 L}{\sqrt{\pi}R_0} \int_0^{+\infty} \exp\left(-\left(\frac{\xi L}{2R_0}\right)^2\right) \frac{I_1\left(\xi\frac{r}{R_0}\right)}{I_1(\xi)} \cos\left(\xi\frac{z+vt}{R_0}\right) d\xi \quad . \qquad (A.23)$$

It should be noted that the integral in (A.20) diverges at low $\xi$, because $I_1(\xi) \sim \xi$ and $I_0(0) = 1$. This is due to the fact that the total current on the surface of the cylinder is not zero in the case of the Gaussian distribution. The divergence disappears if we set the total current, integrated over time, at any point of the surface to be zero.